\documentstyle[psfig,proceedings]{crckapb} 

\def\Msun{\ifmmode{~M_\odot}\else$M_\odot$~\fi}
\def\rsun{\ifmmode{~r_\odot}\else$r_\odot$~\fi}
\def\kms{\ifmmode{$~km\thinspace s$^{-1}~}\else km\thinspace s$^{-1}~$\fi}
\def\ga{\mathrel{\mathchoice {\vcenter{\offinterlineskip\halign{\hfil
$\displaystyle##$\hfil\cr>\cr\noalign{\vskip1.5pt}\sim\cr}}}
{\vcenter{\offinterlineskip\halign{\hfil$\textstyle##$\hfil\cr>\cr
\noalign{\vskip1.0pt}\sim\cr}}}
{\vcenter{\offinterlineskip\halign{\hfil$\scriptstyle##$\hfil\cr>\cr
\noalign{\vskip0.5pt}\sim\cr}}}
{\vcenter{\offinterlineskip\halign{\hfil$\scriptscriptstyle##$\hfil
\cr>\cr\noalign{\vskip0.5pt}\sim\cr}}}}}

\begin{opening}
\title{ARE DAMPED LY$\alpha$ SYSTEMS  \protect\\
       LARGE, GALACTIC DISKS ?}

\author{JESPER SOMMER-LARSEN}
\institute{Theoretical Astrophysics Center\\
           Juliane Maries Vej 30\\
	DK-2100    Copenhagen {\O},  Denmark}

\end{opening}

\runningtitle{ARE DAMPED LY$\alpha$ SYSTEMS LARGE, GALACTIC DISKS ?}

\begin{document}


\begin{abstract}
The hypothesis that the Damped Ly$\alpha$ systems (DLAs) are large, galactic disks
(Milky Way sized) is tested by confronting predictions of models of the formation
and evolution of (large) disk galaxies with observations, in particular the Zinc
abundance distribution with neutral hydrogen column density found for DLAs. A 
pronounced mismatch is found strongly hinting that the majority of DLAs may not
be large, galactic disks.
\end{abstract}

\section{Introduction}
It has been proposed that the Damped Ly$\alpha$ (DLA) systems seen as broad
absorption lines with neutral hydrogen column density $N(HI) \ga$ 2 10$^{20}$ 
cm$^{-2}$ in quasar spectra are large and massive (Milky Way
sized), galactic disks with maximum circular speed $V_c \ga$ 200 km/s (e.g., Prochaska
\& Wolfe 1997, 1998). If true, this is of profound interest in relation to theories
of the formation and evolution of disk galaxies, but it is obviously important to
test this hypothesis in as many ways as possible. In this contribution I present a test
alternative to the ones used by Prochasca \& Wolfe. My test is based on 
fairly elaborate and realistic models of the formation and evolution 
(including chemical evolution) of large and massive disk galaxies, in particular on
comparing the observed distribution of Zinc abundance with neutral hydrogen column
density found for DLAs with model predictions.

In section 2 I briefly describe the models, in section 3 the data, in section 4 the
model predictions are confronted with the data and finally section 5 constitutes a
brief conclusion.

\section{The models}
 
The star-formation rate is assumed to depend on gas surface
density as
\begin{equation}
\frac{d\Sigma_{\star}}{dt} \propto w(\Sigma_{gas}) ~\Sigma_{gas}^N\;\;,
\end{equation}
where N=1-2, and the weight factor
\begin{equation}
w(\Sigma_{gas}) = 1
\end{equation}
for models without a star-formation threshold and 
\begin{eqnarray}
                                        & \simeq 0 ~ , & ~~\Sigma_{gas} \le \Sigma_{TH}
\nonumber \\
w(\Sigma_{gas})  =  \Bigg\{ &
                                  &    \\
                                         & 1 ~ , & ~~\Sigma_{gas} > \Sigma_{TH}
\nonumber
\end{eqnarray}
for models with a star-formation threshold given by (e.g., Kennicutt 1989)
\begin{equation}
\Sigma_{TH} = \alpha ~\frac{\kappa ~c}{3.36 ~G} \simeq 11 ~(\frac{\alpha}{0.7}) 
~(\frac{\kappa}{\kappa_{\odot}}) ~(\frac{c}{6 \mbox{km/s}}) 
~\mbox{M$_{\odot}$/pc$^2$} ~~,
\end{equation}
where $\kappa$ is the epicyclic frequency
\begin{equation}
\kappa = 1.41 ~\frac{V}{R} ~(1 + \frac{R}{V} ~\frac{dV}{dR})^{1/2} ~~,
\end{equation}
$c$ is the velocity dispersion of the gas, $\alpha$ is a dimensionless constant
of the order unity and $\kappa_{\odot}$ is the epicyclic frequency at the
solar distance from the center of the Galaxy. I have adopted $\alpha$ = 0.7 and
$c$ = 6 km/s (Kennicutt 1989). Infall of primordial gas, but no gas out-flows are
included in these models of large and massive disk galaxies. Chemical evolution is calculated
assuming a Scalo (1986) initial mass function (IMF). Oxygen is assumed to be
produced in massive stars only ($M \ga$ 8 M$_{\odot}$) and instantaneously
recycled by Type II supernova (SN) explosions. Iron is assumed to be produced partly
by massive stars ($M \ga$ 8 M$_{\odot}$) and instantaneously recycled by Type II 
SN explosions and partly by less massive stars and recycled by Type Ia SN explosions
with a time-delay of $t_{SNIa}$ = 1 Gyr. Unprocessed gas, locked up in stars, is
assumed to be non-instantaneously recycled.

The proportionality constants in the star-formation
laws (eq. [1]) are determined by applying eq. [1] to the local, solar cylinder and
fitting data on the local gas fraction, G-dwarf distribution etc.
\begin{figure}
\psfig{file=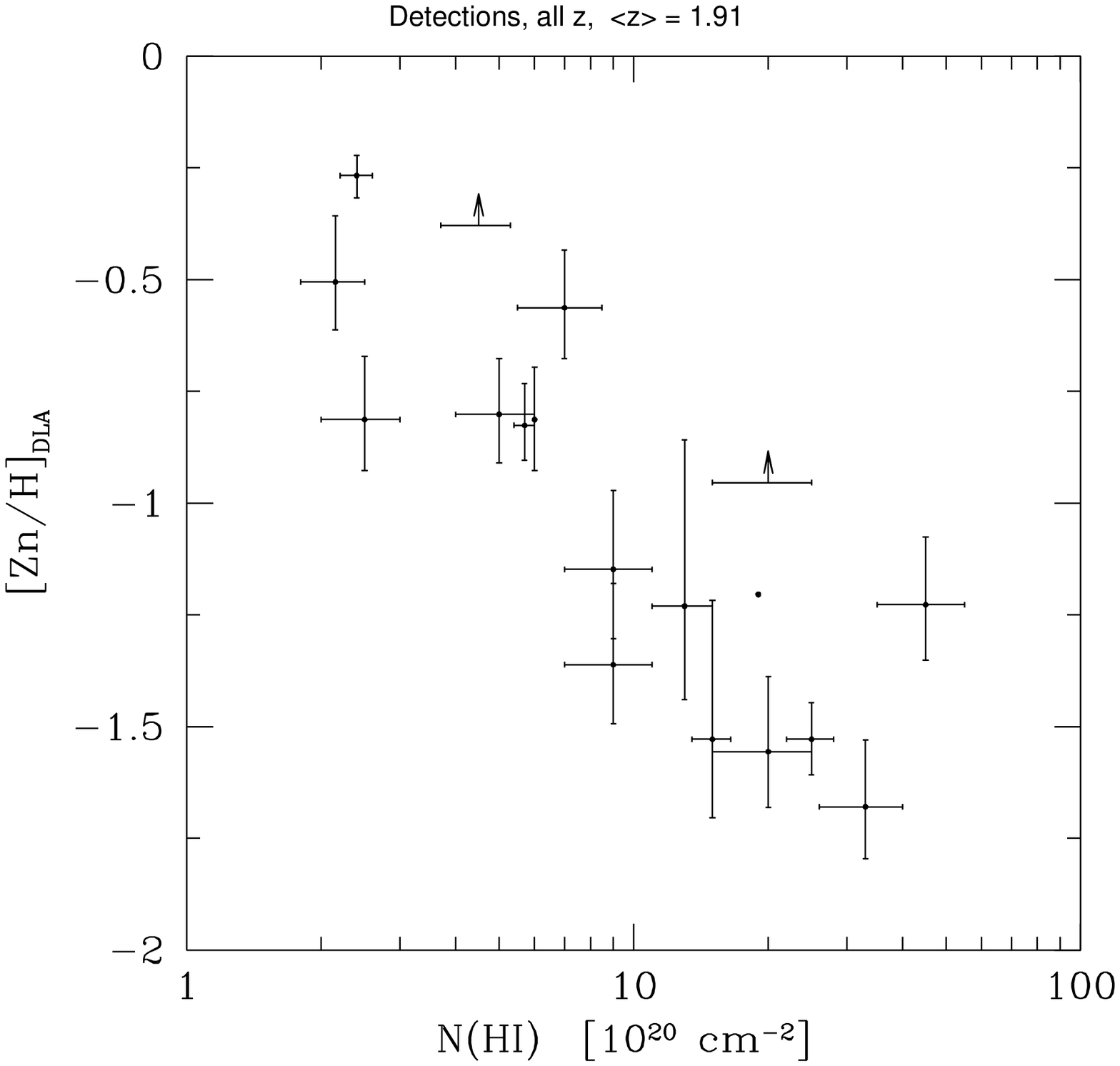,height=12cm,width=12cm}
\caption[]{The observed distribution of $[Zn/H]_{DLA}$ vs. neutral hydrogen column
density $N(HI)$ for the 18 detections in the
combined Pettini et al. sample.}
\end{figure}

\begin{figure}
\psfig{file=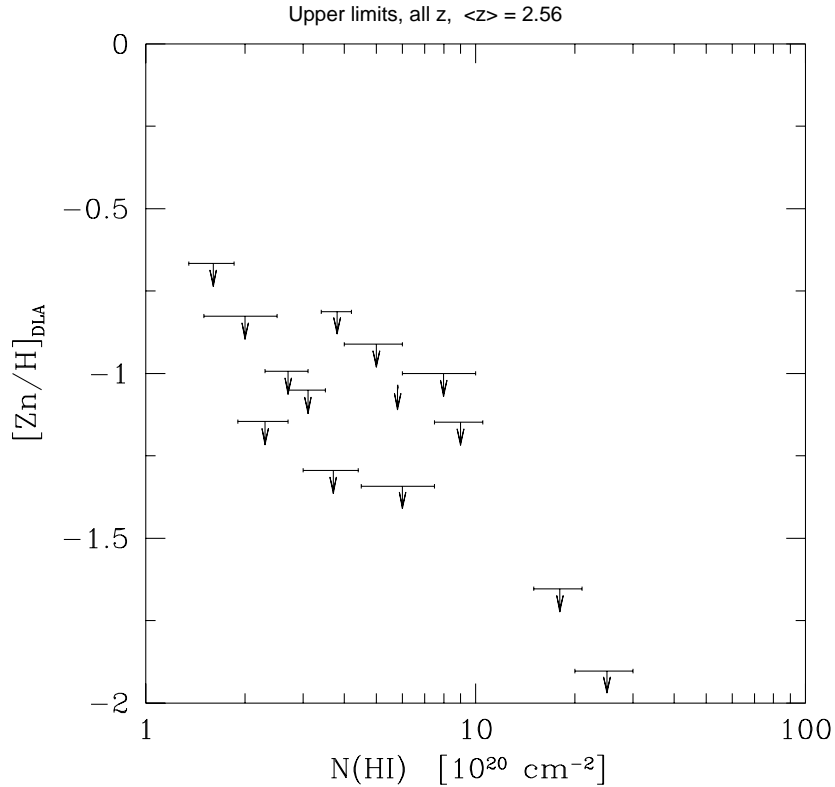,height=12cm,width=12cm}
\caption[]{Same as in Figure 1, but for the 16 upper limits}
\end{figure}

The global properties of the large disks as a function of time (or redshift) are
determined by assuming that an average, spin-parameter $\lambda$ = 0.05, galactic
disk of either early or late type (see below) has an exponentially decreasing stellar profile 
with radius $R$ in the disk (e.g., Freeman 1970) truncated at 4 stellar scale-lengths 
(e.g., van der Kruit 1987) and an exponentially decreasing
oxygen gas abundance profile with radius in the disk (e.g., Zaritsky et al. 1994, 
Garnett et al. 1997). Early type disk galaxies (Sab/Sbc) are assumed to have
an average star-formation history parameter $b$, defined as the ratio of the current to
the past average star-formation rate, of $b \simeq$ 0.33 and late type disks (Scd/Sdm)
to have an average $b \simeq$ 1.0 (Kennicutt et al. 1994). An Ellis (1983) 
present day ($z$=0) morphological
mix of E/S0:Sab/Sbc:Scd/Sdm = 0.28:0.47:0.25 is assumed and the models are 
averaged over spin-parameter $\lambda$. The models thereby cover the full 
range of morphology and global surface brightness of large, galactic disks
and are fully constrained: There are no free parameters left, except for the 
star-formation law index $N$ and the absence or presence of a star-formation
threshold and the conclusions reached in this work turn out to be quite 
insensitive to this for reasonable values of $N$ ($N$=1-2) - see also below.

\section{The data}
Observational data on 34 DLA systems (including Zinc abundances or limits) have
been taken from Pettini et al. (1994, 1997). Zinc is used because it is not or only very
slightly affected by dust depletion. Figure 1 shows $[Zn/H]$ versus neutral 
hydrogen
gas column densities $N(HI)$ for the
18 DLAs with Zinc detected. They have a mean redshift of $<z>$=1.91. Interestingly,
higher Zinc abundance corresponds to {\it lower} $N(HI)$, which is {\it not} to 
be expected
if the DLAs are large, galactic disks, even when effects of dust obscuration of
quasars are taken into account - see below.
Figure 2
shows $[Zn/H]$ versus $N(HI)$ for the 16 DLAs with upper limits on the Zinc
abundance. The mean redshift of these is $<z>$=2.56.

\section{Models versus data}
Figure 3 shows the model predicted $[Fe/H]$ distributions for $N(HI)$ = 2-10 10$^{20}$
\begin{figure}
\psfig{file=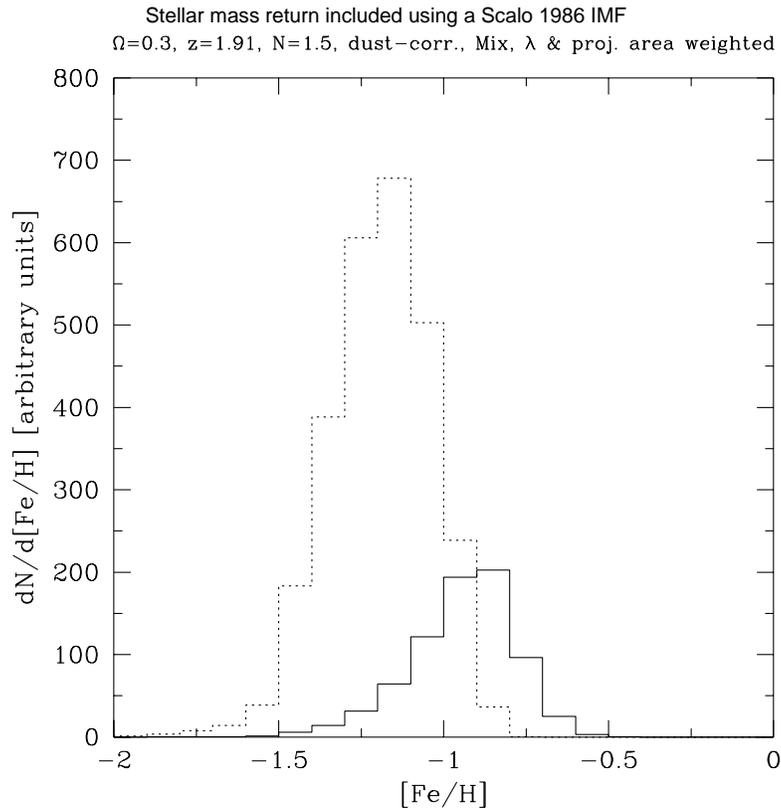,height=12cm,width=12cm}
\caption[]{Predicted DLA abundance distributions at different $N(HI)$ assuming that
DLAs are large, galactic disks similar to the Milky Way: For models with no
star-formation threshold. Dotted line histogram: $N(HI)=2-10 ~10^{20}$ cm$^{-2}$,
Solid line histogram: $N(HI)=10-50 ~10^{20}$ cm$^{-2}$.}
\end{figure}
cm$^{-2}$ and $N(HI)$ = 10-50 10$^{20}$ cm$^{-2}$ for the no star-formation
threshold models with star-formation law index $N$=1.5. The distributions have been
weighted over morphological mix, spin-parameter (and hence effectively global surface
brightness) and projected area and corrected for
the effects of dust obscuration of quasars using the clever methodology developed
by Fall \& Pei (1993).   
Figure 4 shows the same thing, but for the models {\it with}
\begin{figure}
\psfig{file=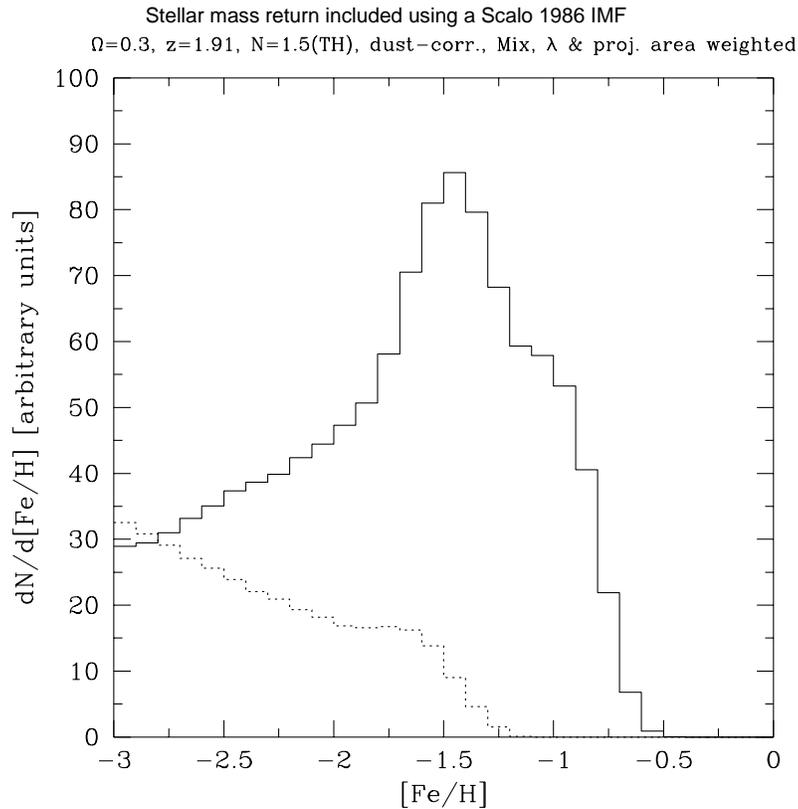,height=12cm,width=12cm}
\caption[]{Same as in Figure 3, but for models with a star-formation threshold}
\end{figure}
a star-formation threshold (and still $N$=1.5). 

Neither of the two sets of model distributions appear to match the data at all
(assuming that $[Zn/H] \simeq
[Fe/H]$ as is the case for Galactic stars of all abundances). More quantitatively,
Komolgorov-Smirnov (KS)
tests show that both sets of model distributions can be ruled out as fits to the data with
more than 99\% confidence (this result is more generally obtained for all values 
$N$=1-2 and also for Wyse \& Silk (1989) type star-formation laws).

The mismatch between the data and the models is further illustrated by Figure 5.
The figure shows the number of predicted DLA systems per redshift
\begin{figure}
\psfig{file=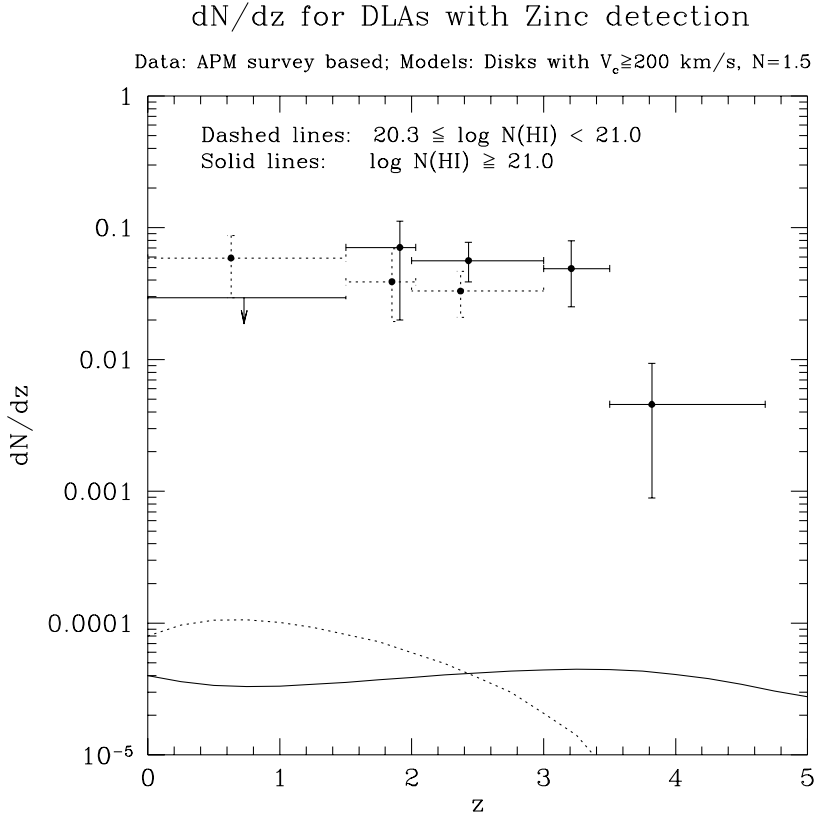,height=12cm,width=12cm}
\caption[]{Predicted number of DLAs per redshift interval for the models with no
star-formation threshold - see text for details. Points with errorbars: Data,
Lines: Models.}
\end{figure}
interval with Zinc detected, $dN/dz$, for $N(HI)$ = 2-10 10$^{20}$ cm$^{-2}$ 
and $N(HI) \ge$ 10$^{21}$ cm$^{-2}$ versus redshift $z$. The models
have been normalized using observational information about the luminosity function
and sizes of present day ($z$=0),  large and massive disk galaxies (with $V_c \ga$ 200 km/s). 
The model
predictions have been weighted and corrected as described above and furthermore
corrected for the effect of a Zinc detection threshold as a function of $N(HI)$.
Figure 5 is for the models with star-formation law index $N$=1.5 and no star-formation
threshold. 
Also shown are APM survey based observational results of Storrie-Lombardi et al.
(1996) corrected by an observationally deduced Zinc detection probability as a function
of redshift $z$ and $N(HI)$ derived from the Pettini et al. (1994, 1997) data. As can be
seen from the figures the two sets of model predictions lie more than two orders of 
magnitude below the observations - the situation is even worse for the models {\it
with} a star-formation threshold (not shown).

\section{Conclusion}
The analysis presented above suggests that the DLA systems may not in general be 
accounted for by large
and massive disk galaxies with $V_c \ga$ 200 km/s, not even when account is taken
for the existence of some very large and low surface brightness disk galaxies 
(physically: high $\lambda$ galactic disks). The true situation is most likely
quite complicated,
but it is tempting to associate the DLAs at low redshift with the large population of
faint, blue dwarfs and at higher redshift with the large number of smaller galaxies
expected at such redshifts ($z \sim$ 2-4) in hierarchical galaxy formation scenarios
(e.g., Kauffmann 1996, Mo et al. 1998). DLAs with indications of large internal
velocities ($\sim$ 200 km/s) may, at least partly, be dynamically unrelaxed systems
(e.g., Haehnelt et al. 1998).

\section*{Acknowledgements}
I have benefited from discussions with Mike Edmunds, Don Garnett, Bernard 
Jones, Bernard Pagel and Max Pettini.

{}

\end{document}